\documentclass[12pt]{article}
\begin{document}
\title{Puzzles of Large Scale Structure and Gravitation}
\author{B.G. Sidharth\\
International Institute for Applicable Mathematics \& Information Sciences\\
Hyderabad (India) \& Udine (Italy)\\
B.M. Birla Science Centre, Adarsh Nagar, Hyderabad - 500 063 (India)}
\date{}
\maketitle
\begin{abstract}
We consider the puzzle of cosmic voids bounded by two dimensional structures of galactic clusters as also a puzzle pointed out by Weinberg: How can the mass of a typical elementary particle depend on a cosmic parameter like the Hubble constant? An answer to the first puzzle is proposed in terms of ``Scaled'' Quantum Mechanical like behaviour which appears at large scales. The second puzzle can be answered by showing that the Gravitational mass of an elementary particle has a Machian character.
\end{abstract}
\section{Large Scale Structures}
Our view of the universe has been continuously evolving over the centuries. Thus Newton's universe was one in which the stars were the building blocks. These building blocks were stationary in the universe. After about two centuries this view underwent a transformation, with the discovery in the early twentieth century by Hubble that the so called galactic nebulae were star systems or galaxies, each containing something like a hundred thousand million stars, and these galaxies themselves being at distances far far beyond those of stars. The building blocks were now the galaxies. Then the Red Shift studies of galaxies by V.M. Slipher showed that the galaxies were all rushing outwards. Thus was born the precursor of what has come to be known as the standard Big Bang Cosmology. Soon it was realized that there were clusters of galaxies which would more correctly qualify as the building blocks, and then super clusters. Within clusters and super clusters, there could be departures from the Hubble velocity distance law. The law therefore represented something happening at a very large scale.\\
A further development that came about in the 1980s threw up a dramatically different scenario. The very large clusters of galaxies seemed to lie on bubble or balloon like sheets, there being voids or very thinly populated regions in the interiors. These voids would have dimensions of the order of a hundred million light years \cite{peebles,cornell,scam,long}. This has been a puzzle thrown up in the late twentieth century: Exactly why do we have the voids and why do we have polymer like two dimensional structures on the surfaces of these voids? The puzzle is compounded by the fact that given the dispersion velocities of the galaxies of the order of a thousand kilometers per second, it would still take periods of time greater than the age of the universe, of the order of 13 billion years, for them to move out of an otherwise uniform distribution, leaving voids in their wake. An interesting suggestion was that the galaxies consisting of ordinary matter were floating on the ``voids'' which are actually made up of dark matter. In any case, latest developments have marginalized dark matter in favour of dark energy.\\
One of the few explanations for this large scale structure of the universe has been the pancake model of Zeldovich \cite{salam}. Essentially according to this model, much of the matter of the universe was in the form of a thin pancake which broke up into pieces, the pieces then forming the clusters of galaxies and galaxies, which have inherited the two dimensional character. Indeed studies have suggested this two dimensional character \cite{bgspopova}.\\
In the above context we consider the model of ``Scaled'' Quantum Effects \cite{csf,csf2,csf3,cu,carn,Agnese}. To sum up the main results: It is argued that the structures of the unvierse at different scales mimic Brownian effects, which again lead to a Quantum behavior with different ``Scaled'' Planck constants. Thus we have,  
\begin{equation}
R \approx l_1 \sqrt{N_1}\label{eaa5}
\end{equation}
\begin{equation}
R \approx l_2 \sqrt{N_2}\label{eaa6}
\end{equation}
\begin{equation}
l_2 \approx l_3 \sqrt{N_3}\label{eaa7}
\end{equation}
\begin{equation}
R \sim l \sqrt{N}\label{eaa8}
\end{equation}
and a similar relation for the KBO (Kuiper Belt objects)
\begin{equation}
L \sim l_4 \sqrt{N_4}\label{eaa9}
\end{equation}
where $N_1 \sim 10^6$ is the number of superclusters in the universe,
$l_1 \sim 10^{25}cms$ is a typical supercluster size $N_2 \sim
10^{11}$ is the number of galaxies in the universe and $l_2 \sim
10^{23}cms$ is the typical size of a galaxy, $l_3 \sim 1$ light years
is a typical distance between stars and $N_3 \sim 10^{11}$ is the
number of stars in a galaxy, $R$ being the radius of the universe
$\sim 10^{28}cms, N \sim 10^{80}$ is the number of elementary
particles, typically pions in the universe and $l$ is the pion Compton
wavelength and $N_4 \sim 10^{10}, l_4 \sim 10^5 cm$, is the dimension
of a typical KBO (with mass $10^{19}gm$ and $L$ the width of the
Kuiper Belt $\sim 10^{10}cm$ (Cf.ref.\cite{cu})).\\
These in turn lead to the ``Scaled'' Planck constants
\begin{equation}
h_1 \sim 10^{93}\label{eaa1}
\end{equation}
for super clusters;
\begin{equation}
h_2 \sim 10^{74}\label{eaa2}
\end{equation}
for galaxies and
\begin{equation}
h_3 \sim 10^{54}\label{eaa3}
\end{equation}
for stars. And
\begin{equation}
h_4 \sim 10^{34}\label{eaa4}
\end{equation}
for Kuiper Belt objects.\\
Infact Equations (\ref{eaa5}) to (\ref{eaa9}) correspond to the empirically well known Weyl-Eddington formula. It was argued that they express Random Walk effects in a Nelson like approach \cite{psp,psu,nottale}. The origin of Equations (\ref{eaa5}) to (\ref{eaa9}) is, rather than being empirical, due to gravitational orbits and the conservation of angular momentum viz.,
\begin{equation}
\frac{GM}{L} \sim v^2, M v L = H\label{eaa11}
\end{equation}
where $L,M,v$ represent typical length
(or dispersion in length), mass and velocities at that scale and $H$
denotes the scaled Planck constant.\\
We can further deduce on the above basis that 
\begin{equation}
M = \left(\frac{\hat{H}H^2}{Gv}\right)^{\frac{1}{3}}\label{ex}
\end{equation}
where 
$$\hat{H} = \frac{v}{L}$$
is the analogue of the Hubble constant. It may be pointed out that (\ref{ex}) is itself the analogue of a relation of identical form, viz.,
\begin{equation}
m \approx \left(\frac{H'\hbar^2}{Gc}\right)^{\frac{1}{3}}\label{e1}
\end{equation}
where $m$ is the mass of an elementary particle like the pion, $H'$ the Hubble constant and $\hbar,G$ and $c$ have their usual meaning. Infact there is a complete parallel with Quantum theory. For example we have for the universe at large
\begin{equation}
R = \frac{h_1}{Mc}\label{eaa10}
\end{equation}
The above considerations lead via the diffusion process to the Schrodinger like equation \cite{nottale,univfl},
\begin{equation}
h_\imath \frac{\partial \psi}{\partial t} + \frac{h^2_\imath}{2m}
\nabla^2 \psi = 0\label{eaa14}
\end{equation}
for different $h_\imath$ given by equations (\ref{eaa1}) to (\ref{eaa4}).\\
Before proceeding further we may point out that (\ref{eaa5}) to (\ref{eaa9}) already indicate the two dimensionality referred to above. Infact alternatively, the theory is modelled on a phase transition viz., the Landau-Ginzburg theory applied to an equation like (\ref{eaa14}). (Such a phase transitioin would also explain what the movement of galaxies under normal circumstances cannot, that is the large size of the voids.) Infact to put it briefly, under such phase transitions we have equations like 
$$(\bar Q) = |t|^\beta , (\bar \xi) = |t|^{-\nu}$$ 
Whence
\begin{equation}
\bar Q^\nu = \bar \xi^\beta\label{e18a}
\end{equation}
where typically $\nu \approx 2\beta$ and $\bar{Q} \sim \frac{1}{\sqrt{N}}$ is the reduced order parameter and $\bar{\xi}$ is the reduced correlation length. We then have,
\begin{equation}
\bar Q \sim \frac{1}{\sqrt{N}}, \bar \xi = (l/R)^2\label{e19a}
\end{equation}
Whence on using (\ref{e18a}) we recover the Weyl-Eddington like equations (\ref{eaa5}) to (\ref{eaa9}). This is yet another derivation.\\
Let us see if the above model can give an estimate for the size of the voids. Infact we have to revert to the Schrodinger like equation (\ref{eaa14}) with a hydrogen like atom, except that $GM^2$ replaces $Ze^2$ of the hydrogen atom. Let us consider the hydrogen like wave functions at a scale of galaxies with $h_\imath$ replaced by $h_2$ given above. The radial part of a typical wave functioin would be given by \cite{paul}
\begin{equation}
\psi_l = \left\{ \left(6-6 \rho + \rho^2\right) e^{-\rho^{\frac{1}{2}}}\right\}\label{ez}
\end{equation}
It is easy to verify that the expression (\ref{ez}) is a maximum with $\rho$ given by,
\begin{equation}
\rho = \frac{2Ze^2r}{h^2_2} M \to \frac{2GM^3 r}{h^2_2} \sim 10\label{ez'}
\end{equation}
Infact (\ref{ez'}) gives us back, the Weinberg like formula (\ref{ex}) encountered earlier. We thus have from (\ref{ez'}) after a simple calculation and feeding in the values for $h_2$ and $M \sim 10^{44}gm$, that $r \sim 100$ light years, exactly as required.\\
The point is that at radial distances like $r$ given above, there would be a greater concentration of galaxies while within this value of $r$ the distribution would be comparatively sparse. We must also remember that there is the angular part of the ``wave function'', which means that each value of $r$ really corresponds to a spherical shell.\\
Thus it is a consequence of Scaled Quantum Effects arising due to the gravitational forces (\ref{eaa11}) that lead to the bubble and void structure of the universe. Similar arguments could be put forth for the pancake structure of galaxies themselves.
\section{The Puzzle of Gravitation}
There is a well known relation, some times referred to as the Weinberg formula, which is (\ref{e1}).
 Equation (\ref{e1}) has been considered to be a purely accidental relation. However as Weinberg notes \cite{wein}  ``...it should be noted that the particular combination of $\hbar , H, G$, and $c$ appearing (in the formula) is very much closer to a typical elementary particle mass than other random combinations of these quantities; for instance, from $\hbar , G$, and $c$ alone one can form a single quantity $(\hbar c/G)^{1/2}$ with the dimensions of a mass, but this has the value $1.22 \times 10^{22} MeV/c^2$, more than a typical particle mass by about $20$ orders of magnitude!\\
``In considering the possible interpretations (of the formula), one should be careful to distinguish it from other numerical ``coincidences''... In contrast, (the formula) relates a single cosmological parameter, $H$, to the fundamental constants $\hbar , G, c$ and $m$, and is so far unexplained.''\\
We will now argue that (\ref{e1}) represents a Machian or holistic effect.\\
Let us consider the gravitational self interaction of a particle (Cf. also ref.\cite{bgspopova2}). Our starting point is the action functional
$$S = -(8\pi G)^{-1} \int d^4 x \phi \Delta^2 \phi + \int d^4 x \Psi^* \left(\imath \hbar \frac{\partial \Psi}{\partial t} + \frac{\hbar^2}{2m} \Delta^2 \Psi - m \phi \Psi\right)$$
where $\phi$ is some potential whose nature is not as yet specified, $G$ being some coupling constant. The extremum conditions of action with respect to $\Psi^*$ and $\Psi$ lead to the Schrodinger equation with the interaction potential $\phi$:
\begin{equation}
\imath \hbar \frac{\partial \Psi}{\partial t} = -\frac{\hbar^2}{2m} \Delta^2 \Psi + m \phi \Psi\label{ea1}
\end{equation}
and to the Poisson equation for the potential itself
\begin{equation}
\Delta^2 \phi = 4 \pi Gm\Psi^* \Psi\label{ea2}
\end{equation}
Thus, the equations (\ref{ea1}) and (\ref{ea2}) describe a self-interacting particle. It is well known that an exact solution to (\ref{ea2}) is given by
\begin{equation}
\phi (\vec {r},t) = -G \int_{\Omega} d \Omega (\vec{r}) \frac{\rho (\vec{r},t)}{|\vec{r} - \vec{r}'|},\label{ea3}
\end{equation}
where $\Omega$ is the three dimensional region which confines the particle, and we have defined
\begin{equation}
\rho (\vec{r},t) = m \Psi^* (\vec{r},t) \Psi (\vec{r},t)\label{ea4}
\end{equation}
From (\ref{ea3}), we can immediately see that for distances far outside the region $\Omega$, that is $|\vec{r}| < < |\vec{r}'|$, the potential $\phi$ has the form
\begin{equation}
\phi \approx \frac{GM}{r},\label{ea5}
\end{equation}
where $r = |\vec{r}|$, and we have defined $M$ as,
\begin{equation}
M = \int_{\Omega} d\Omega (\vec{r}) \rho (\vec{r},t) = m \int_{\Omega} d\Omega (\vec{r})\Psi^* (\vec{r},t) \Psi (\vec{r},t)\label{ea6}
\end{equation}
The integral on the right hand side of (\ref{ea6}) is conserved in time due to (\ref{ea1}):
$$\frac{\partial}{\partial t} \int_{\Omega} d\Omega (\vec{r}) \Psi^* (\vec{r},t) \Psi (\vec{r},t) = 0$$
Thus the quantity $M$ is constant, and we can interpret (\ref{ea5}) and (\ref{ea6}) as follows. The attractive potential (\ref{ea5}) is now the classical gravitational potential, $M$ is the gravitational mass, $G$ being the gravitational constant. If we prescribe the unit value to the above conserved functional and interpret it as the norm square, $I^2$, or the full probability
$$I^2 = \int_{\Omega} d\Omega (\vec{r}) \Psi^* (\vec{r},t) \Psi (\vec{r},t) = 1,$$
then the gravitational mass coincides with the inertial mass,
\begin{equation}
m = M,\label{ea9}
\end{equation}
and the quantity (\ref{ea4}) now can be interpreted as the mass probability denslity. The source term on the right side of (\ref{ea2}) is equal to the particle probability density itself.\\
Now, let us consider the self-consistent problem - the particle in its own potential well. We cannot obtain an exact solution. However, we can approximately describe some features of such a solution. The first assumption will be that we deal only with a spherically symmetric wave function: $\Psi = \Psi (r,t)$ where $r$ is a radial coordinate. Then the mass probability density has the same dependence: $\rho = \rho (r,t)$. It can be easily shown that for any spherical mass distribution, the potential (\ref{ea3}) is reducible to a simple form
\begin{equation}
\phi (r ,t) = G \int^r_0 dr' \frac{m(r' ,t)}{r'^2} - \int^\infty_0 dr' \frac{m(r' ,t)}{r'^2},\label{ea10}
\end{equation}
where we denote
$$m(r,t) = 4 \pi \int^r_0 dr' r'^2 \rho (r' ,t),$$
and $m(r,t)$ is just the mass inside a ball of radius $r$. Certainly, the solution (\ref{ea10}) gives an exact formula (\ref{ea5}) with the mass (\ref{ea9}) for the point mass distribution. Further, we shall use the value $\Phi$ instead of the potential $\phi$:
$$\phi (r,t) = mG\Phi (r,t)$$
This allows us to rewrite (\ref{ea1}) in the form
\begin{equation}
\imath \frac{2m}{\hbar} \frac{\partial}{\partial t} \Psi + \Delta^2 \Psi - \frac{2m^3G}{\hbar^2} \Phi \Psi = 0\label{ea12}
\end{equation}
The coefficient of $\Phi$ in (\ref{ea12}) has the dimensionality of inverse length. Thus, we denote
\begin{equation}
l_G = \frac{\hbar^2}{2m^3G},\label{ea13}
\end{equation}
Equation (\ref{ea13}) is nothing but (\ref{e1}) the Weinberg formula again if we identify $l_G$ with the radius of the Universe and rememeber that,
$$H = \frac{c}{l_G}$$
All this shows that the mass $m$ of an elementary particle is very Machian, rather than being microphysical, if $G$ is microphysical.\\
However in a fluctuational model of cosmology, which correctly predicted in advance a dark energy driven accelerating Universe with a small cosmological constant, (\ref{e1}) and several other, so called Large Number Relations made famous by Dirac \cite{ijmpa,cu,nar} were actually deduced from the theory. In this model it turns out that 
\begin{equation}
G = \frac{lc^2}{m\sqrt{N}}\label{eax}
\end{equation}
where $N \sim 10^{80}$ is the number of elementary particles in the Universe. Indeed (\ref{eax}) was shown to be an alternative form of (\ref{e1}). Moreover it was argued \cite{bgsfpl,univfl} that (\ref{eax}) or equivalently (\ref{e1}) shows up gravitation as a distributional effect over the $N$ elementary particles which constitute the Universe. In this case $G$ would not be a microphysical constant.\\
The following argument throws further light on the above considerations. We identify the inertial energy of a typical elementary particle with its gravitational energy due to all the remaining $N$ elementary particles in the Universe. This gives
$$\frac{GNm^2}{R} = mc^2$$
Using the well known so called Weyl-Eddington formula,
$$R = \sqrt{N}l,$$
which infact was deduced in the cosmological model referred to (Cf.ref.\cite{ijmpa,cu}), the above becomes identical to (\ref{eax}) or (\ref{e1}).\\
It is ofcourse well known that attempts to unify gravitation and electromagnetism, starting with the attempt of Hermann Weyl to present day Quantum Gravity approaches have not been very successful \cite{univfl}. Indeed While Pauli had remarked in this context that one should not try to unite what God had intended to be asunder, as Witten put it, \cite{witten} ``The existence of gravity clashes with our description of the rest of physics by quantum fields''. The above considerations show that gravitation has a different, indeed distributional character. However, (\ref{eax}), on using $e^2 \sim \hbar c$ leads to the well known Large Number Relation (Cf.\cite{wein} and \cite{cu}),
$$\frac{Gm^2}{e^2} \sim \frac{1}{\sqrt{N}} \sim 10^{-40}$$
which infact expresses a connection between the gravitational and electromagnetic coupling constants, which has now been deduced from theory, rather than being an ad hoc accidental relation.

\end{document}